\documentclass[12pt,a4paper]{amsart}
\usepackage[T1]{fontenc}
\usepackage{lmodern}
\usepackage[margin=25mm]{geometry}
\usepackage{amssymb,booktabs,microtype}
\usepackage[hidelinks]{hyperref}

\hypersetup{
  pdftitle={A new upper bound for A(19,6,8)},
  pdfauthor={Jiang Chang}
}
\newtheorem{theorem}{Theorem}
\newcommand{\cC}{\mathcal{C}}
\newcommand{\R}{\mathbb{R}}
\newcommand{\Ustar}{U^{*}}
\DeclareMathOperator{\wt}{wt}
\DeclareMathOperator{\tr}{tr}

\title{A new upper bound for $A(19,6,8)$}
\author{Jiang Chang}
\address{College of Computer Science and Software Engineering,
  Hohai University, Nanjing, China}
\email{science614@163.com}
\thanks{ORCID: \href{https://orcid.org/0009-0006-8310-632X}{0009-0006-8310-632X}}
\date{30 October 2026}

\begin{document}

\begin{abstract}
We compute Polak's bound $A_4(19,6,8)$ for binary constant weight codes,
obtaining the numerical value $673.04001$. An exact rational certificate
gives $A_4(19,6,8)\le 673.0400143064802474039375132665301315202504$,
and hence $A(19,6,8)\le 673$. This improves the upper bound $693$ listed
in Brouwer's table. The certificate is verified by a separate program
using exact rational arithmetic.
\end{abstract}

\maketitle

\section{Introduction and main result}

Let $A(n,d,w)$ be the maximum size of a binary code of length $n$,
constant weight $w$, and minimum Hamming distance at least $d$.
We compute Polak's semidefinite programming bound $A_4(n,d,w)$
\cite{Polak} at $(n,d,w)=(19,6,8)$. This bound adapts the quadruple
bound of Gijswijt, Mittelmann, and Schrijver \cite{GMS} to constant
weight codes.

Brouwer's table \cite{Brouwer}, accessed on 30 October 2026, lists
$408\le A(19,6,8)\le 693$, with the upper bound attributed to Polak.
In Polak's Table~2, asterisks distinguish $A_4$ values from $B_4$
values: the entries for $(17,6,7)$, $(18,6,7)$, and $(19,6,7)$ are
$206^{*}$, $312^{*}$, and $463^{*}$, respectively. The entry for
$(19,6,8)$ is the unmarked value $693$, obtained by rounding down
$B_4(19,6,8)$.

\begin{theorem}\label{thm:main}
We have $A(19,6,8)\le A_4(19,6,8)\le\Ustar$, where
\begin{align*}
\Ustar
&=\frac{841300017883100309254921891583162664400313}
        {1250000000000000000000000000000000000000}\\[6pt]
&=673.0400143064802474039375132665301315202504.
\end{align*}
In particular, $A(19,6,8)\le 673$.
\end{theorem}

\section{The bounds \texorpdfstring{$A_4$ and $B_4$}{A4 and B4}}

We recall Polak's definitions \cite[Section~1.1]{Polak}.
Let $F=\{u\in\{0,1\}^n:\wt(u)=w\}$ and
$\cC_j=\{C\subseteq F:|C|\le j\}$. For $D\in\cC_j$, set
\[
\cC_j(D)=\{C\in\cC_j:D\subseteq C,\ |D|+2|C\setminus D|\le j\},
\qquad
M_{j,D}(x)=\bigl(x(C\cup C')\bigr)_{C,C'\in\cC_j(D)}.
\]
The bound $A_k(n,d,w)$ is the maximum of $\sum_{u\in F}x(\{u\})$
over $x:\cC_k\to\R$ satisfying $x(\varnothing)=1$, $x(S)=0$
whenever $S$ contains two distinct words at distance less than $d$,
and $M_{k,D}(x)\succeq0$ for every $D\in\cC_k$.
For $k\ge3$, the relaxation $B_k(n,d,w)$ has the same variables,
objective, and normalization and distance conditions, with matrix
constraints
\[
\begin{cases}
M_{k-1,D}(x)\succeq0,&D\in\cC_{k-1},\ |D|<2,\\
M_{k,D}(x)\succeq0,&D\in\cC_k,\ |D|\ge2.
\end{cases}
\]
Polak's Proposition~1.1 gives
\begin{equation}\label{eq:hierarchy}
A_{k-1}(n,d,w)\ge B_k(n,d,w)\ge A_k(n,d,w)\ge A(n,d,w).
\end{equation}
For the last inequality, a code $C\subseteq F$ of minimum distance
at least $d$ gives a feasible function $x(S)=\mathbf{1}_{S\subseteq C}$
with objective value $|C|$.
For $A_4$, the matrices with $|D|=1$ or $3$ are principal
submatrices of matrices with $|D|=0$ or $2$, respectively;
the $|D|=4$ constraints are scalar nonnegativity conditions.
Averaging over coordinate permutations allows $x$ to be constant
on each orbit, as in Polak's symmetry reduction.

\section{The program for \texorpdfstring{$(19,6,8)$}{(19,6,8)}}

Our implementation of Polak's symmetry reduction gives a program
with $8603$ normalized orbit variables $z_v$. The singleton orbit
variable $z_{\bullet}$ is normalized to equal the code size, and
is the objective. The normalized variables satisfy
\begin{equation}\label{eq:box}
0\le z_v\le\binom{19}{8}=75582.
\end{equation}
These bounds hold at every feasible point: the lower bounds are
constraints of the program, and the upper bounds follow from
$2\times2$ principal minors of the matrices $M_{4,D}$ together with
$x(\varnothing)=1$.
Before compression, the program has $9126$ blocks:
\begin{center}
\begin{tabular}{lrr}
\toprule
Source & Number of blocks & Largest order\\
\midrule
$M_{4,\varnothing}$ & $92$ & $130$\\
$M_{4,D}$ with $|D|=2$ & $431$ & $80$\\
Scalar nonnegativity & $8603$ & $1$\\
\bottomrule
\end{tabular}
\end{center}
Exact compression, supported by $523$ block identities verified
exactly, gives $9123$ blocks of largest order $79$.
An exact invertible diagonal congruence scaling is applied to
every block. The final program contains 59,471,389 nonzero
coefficients.

\section{Numerical solution and exact certificate}

MOSEK~11.2.7 \cite{MOSEK} was run with its interior-point method
and $4$ threads on a Linux server with $16$ logical CPUs and
$256$~GB of memory. It performed $88$ iterations and reported
solution status \texttt{optimal}, terminating because of slow
progress near the optimum. Solver time was about $7935$ seconds,
with peak memory about $39$~GB. The numerical value is
$A_4(19,6,8)\approx673.04001$.

Starting from MOSEK's floating-point dual candidate, we restore
the original coordinates using the verified exact scaling records.
Homogeneous nonnegative scalar multipliers are set to zero.
The remaining blocks are symmetrized and balanced; a numerical
eigendecomposition, with negative eigenvalues clipped to zero,
supplies Gram factors. Each factor entry is rounded to the nearest
multiple of $10^{-20}$, and rational Gram matrices are formed
after undoing the balancing. Positive semidefiniteness is checked
by exact rational $LDL^{\mathsf T}$ factorization with nonnegative
diagonal entries. All $9123$ blocks pass this check.

\begin{proof}[Proof of Theorem~\ref{thm:main}]
Let $P_b(z)\succeq0$ denote the blocks in the exact compressed
program, and let $Y_b\succeq0$ be the rational certificate blocks.
With $\langle Y,P\rangle=\tr(YP)$, exact recombination of all
coefficients gives a constant $r_0$ and residual coefficients
$r_v$ satisfying
\[
z_{\bullet}+\sum_b\langle Y_b,P_b(z)\rangle
=r_0+\sum_v r_vz_v.
\]
Positive semidefiniteness and \eqref{eq:box} therefore give
\begin{equation}\label{eq:certificate}
z_{\bullet}\le r_0+75582\sum_v\max(r_v,0).
\end{equation}
For this certificate, the compensation term is exactly zero and
$r_0=\Ustar$. Taking the maximum over feasible $z$, then using
\eqref{eq:hierarchy} and the integrality of $A(19,6,8)$, proves
the theorem.
\end{proof}

A separate verifier uses exact rational arithmetic from Python's
standard-library \texttt{fractions} module. It checks the model
hashes, reconstructs every $LDL^{\mathsf T}$ matrix, streams all
59,471,389 model coefficients, and recomputes every residual
and the bound, reproducing $\Ustar$ exactly. Certificate generation
took about $223$ seconds, and independent verification about
$892$ seconds.

\section{Consistency checks}

Two codes with parameters $(19,6,8)$, of sizes $227$ and $221$,
were found by randomized greedy search. In exact arithmetic,
the vectors induced by these codes satisfy all $9126$
semidefinite constraints of the uncompressed program.
With the same program generator, we obtain
$A_4(18,6,7)\approx312.76207$ and
$A_4(19,6,7)\approx463.67868$; their integer parts are $312$
and $463$, the entries marked as $A_4$ in Polak's Table~2
\cite{Polak}.

\subsection*{Data availability}
The certificate, the semidefinite program and the verification
program are available from the author upon request.

\subsection*{Use of AI tools}
The computer code used in this work and the text of this note were
prepared with the assistance of AI models (GPT and Claude). The
author has checked all content and takes full responsibility for it.

\end{document}